\documentclass[aps,prx,twocolumn,floats,showpacs,superscriptaddress,nofootinbib,longbibliography]{revtex4-2}
\usepackage{graphicx,epsfig}
\usepackage{times,bbm}
\usepackage{graphics,dcolumn,bm,float}
\usepackage{amssymb,amsmath,rotate,color,amsfonts}
\usepackage[title,titletoc,toc]{appendix}
\usepackage{mathtools}
\usepackage{booktabs}
\usepackage{tcolorbox}
\usepackage[pagebackref=false,colorlinks,linkcolor=magenta,citecolor=blue,urlcolor=magenta]{hyperref}
\usepackage{wrapfig}
\usepackage{lipsum}
\usepackage{mwe}
\usepackage[mathlines]{lineno}
\usepackage[mathscr]{euscript}
\usepackage{hyperref}
\usepackage{breqn}
\usepackage{bbold}
\usepackage{pgfplots}
\usepackage{float}
\usepackage{tkz-euclide}
\usepackage{braket}
\usepackage{physics}
\usepackage[caption=false]{subfig}
\usepackage[export]{adjustbox}
\usepackage{tikz}
\usepackage{slashed}
\usetikzlibrary{through,calc}
\usetikzlibrary{positioning}
\usepackage{mathrsfs,eufrak}
\usepackage{xcolor}

\newcommand{\be}{\begin{equation}}
\newcommand{\ee}{\end{equation}}

\newcommand{\bea}{\begin{eqnarray}}
\newcommand{\eea}{\end{eqnarray}}

\newcommand{\bmt}{\left[\begin{matrix}}
\newcommand{\emt}{\end{matrix}\right]}

\begin{document}
\title{From Chern to Winding: Topological Invariant Correspondence in the Reduced Haldane Model}
\author{Ghassan Al-Mahmood}
\affiliation{Faculty of Physics$,$ University of Isfahan$,$ Isfahan 81746-73441$,$ Iran}
\author{Mohsen Amini}\email{email@msn.amini@sci.ui.ac.ir}
\affiliation{Faculty of Physics$,$ University of Isfahan$,$ Isfahan 81746-73441$,$ Iran}
\author{Ebrahim Ghanbari-Adivi}
\affiliation{Faculty of Physics$,$ University of Isfahan$,$ Isfahan 81746-73441$,$ Iran}
\author{Morteza Soltani}
\affiliation{Faculty of Physics$,$ University of Isfahan$,$ Isfahan 81746-73441$,$ Iran}

\begin{abstract}

We present an exact analytical investigation of the topological properties and edge states of the Haldane model defined on a honeycomb lattice with zigzag edges. By exploiting translational symmetry along the ribbon direction, we perform a dimensional reduction that maps the two-dimensional model into a family of effective one-dimensional systems parametrized by the crystal momentum $k_x$. Each resulting one-dimensional Hamiltonian corresponds to an extended Su-Schrieffer-Heeger (SSH) model with momentum-dependent hoppings and onsite potentials. We introduce a natural rotated basis in which the Hamiltonian becomes planar and the winding number ($\nu$) is directly computable, providing a clear topological characterization of the reduced model. 
This framework enables us to derive closed-form expressions for the edge-state wavefunctions and their dispersion relations across the full Brillouin zone. 
We show that the $\nu$ exactly reproduces the Chern number of the parent model in the topologically nontrivial phase and allows for an exact characterization of the edge modes. Analytical expressions for the edge-state wavefunctions and their dispersion relations are derived without requiring perturbative methods. Our analysis further reveals the critical momentum $ k_c $ where edge states traverse the bulk energy gap, and establishes precise conditions for the topological phase transition.
In contrast to earlier models, such as plaquette-based tight-binding reductions, our method reveals hidden geometric symmetries in the extended SSH structure that are essential for understanding the topological behavior of systems with long-range hopping.
Our findings offer new insight into the topological features of zigzag nanoribbons and establish a robust framework for analyzing analogous systems. This dimensional reduction perspective may also shed light on other long-range SSH systems and symmetry-protected topological phases in higher-dimensional lattices.

\end{abstract}
\maketitle
\section{Introduction~\label{Sec01}}
Topological phases~\cite{RMP1, RMP2, RMP3, HasanKane2010} have revolutionized condensed‐matter physics by revealing that electronic systems can host quantized responses and robust boundary modes, even in the absence of conventional symmetry‐breaking order. Two‐dimensional (2D) time‐reversal‐breaking systems were among the first to exhibit nontrivial topology, with the integer quantum Hall effect demonstrating that the Hall conductance is determined by the Chern number—an integer invariant computed from the Berry curvature of Bloch bands—and measured directly via precisely quantized plateaus in transport experiments~\cite{Thouless1982}. Crucially, the bulk band structure in both topologically trivial and nontrivial phases remains gapped and indistinguishable by local probes, rendering edge‐state measurements essential for detecting the underlying topology~\cite{HasanKane2010}. This bulk–boundary correspondence, which links a nonzero bulk invariant to the existence of conducting surface or edge modes, underpins the robustness of topological insulators and superconductors against perturbations and disorder. A comprehensive classification of these phases—encompassing $\mathbb{Z}$ and $\mathbb{Z}_2$ invariants across various symmetry classes—was established through the tenfold way, systematically organizing possible topological states according to their symmetries and spatial dimensions~\cite{Schnyder2008,Kitaev2009, RMP4}. Together, these insights have not only deepened our theoretical understanding but also guided the experimental discovery of new materials and device concepts based on symmetry‐protected topological order.

The Haldane model, proposed in 1988, stands as the canonical example of a Chern insulator: a system with zero net magnetic flux yet broken time‐reversal symmetry achieved via complex next‐nearest‐neighbor hoppings on a honeycomb lattice~\cite{Haldane1988}. Its tight‐binding Hamiltonian incorporates a staggered sublattice potential alongside a complex phase $\varphi$ acquired in the second‐neighbor hopping, leading to a quantized Hall conductance in the absence of Landau levels. Despite its conceptual simplicity, the Haldane model captures the essential features of more complex materials exhibiting the quantum anomalous Hall effect~\cite{Chang2013}. In finite geometries—such as nanoribbons with zigzag or armchair edges—this model predicts unidirectional edge states whose dispersion and localization can be calculated exactly, offering a versatile platform for both theoretical exploration and potential device applications.

In finite‐width geometries such as nanoribbons, the bulk–boundary correspondence of the Haldane model becomes manifest through the emergence of unidirectional edge channels confined to the zigzag or armchair terminations~\cite{Hatsugai1993, Rahmati, Sadeghizadeh}. In particular, zigzag edges on the honeycomb lattice support chiral modes that traverse the bulk gap, whose properties  can be tuned by the complex hopping phase $\varphi$ and the sublattice‐staggered potential~\cite{R1, R2, R3, R4, R5, R6, R7}. These edge states are topologically protected against backscattering, giving rise to quantized conductance in ideal ribbons. Nanoribbon geometries therefore offer an experimentally accessible platform to probe topological invariants via transport measurements and scanning tunneling spectroscopy, as they allow direct visualization and manipulation of edge‐mode dispersion and localization profiles~\cite{Lee2015}.

A central challenge in analyzing nanoribbon edge modes lies in connecting their 2D topology to tractable 1D descriptions. The basic idea of dimensional reduction—fixing the transverse Bloch momentum and mapping the parent 2D Hamiltonian onto an effective 1D chain—has proven invaluable in numerous contexts, from quantum Hall edges~\cite{Hatsugai1993} to superconducting heterostructures~\cite{Zhang2023}. In this work, we exploit the translational invariance along the ribbon direction to fix the Bloch momentum $k_x$, thereby mapping the Haldane Hamiltonian at each $k_x$ onto an extended SSH model parameterized by $k_x$. This procedure transforms the Chern number of the 2D system into a momentum‐resolved winding number ($\nu$) of the effective chain, which can be computed exactly and visualized as the loop traced by the Bloch vector in a plane. Recent applications of similar dimensional‐reduction techniques include analyses of higher‐order topological phases~\cite{Ad8378} and moiré superlattices~\cite{NatureS41535}, demonstrating the broad utility of this approach. Our method not only yields closed‐form expressions for edge‐state energies, wave functions, and critical momenta $k_c$, but also provides a fully nonperturbative derivation of the bulk–boundary correspondence in chiral insulators.  

Despite significant progress, earlier studies of Haldane‐ribbon edge states have relied on perturbative treatments or numerical diagonalization, without fully elucidating their topological origin. For instance, our prior work~\cite{Rahmati} derived analytical edge‐state wavefunctions via a series expansion but did not connect these solutions to a bulk topological invariant. Similarly, other approaches have calculated edge dispersions numerically but lacked a clear framework for predicting critical momenta or understanding the impact of additional symmetry‐breaking terms~\cite{Lee2015}. More recently, the emergence of topological bound states in Haldane‐model zigzag nanoribbons was demonstrated numerically~\cite{Ad8378}, highlighting novel localized modes beyond the conventional chiral edges; however, that study did not present a systematic mapping to a one‐dimensional winding number nor analytic expressions for edge‐state energies or phase boundaries. In contrast, our dimensional‐reduction method provides a unified, nonperturbative pathway: by mapping each fixed $k_x$ slice of the Haldane model onto an extended SSH chain, we derive exact formulas for the winding number, edge‐state energies, wavefunctions, and the band‐edge momentum $k_c$, as well as precise topological phase boundaries across all $(k_x,M)$. This comprehensive analytical framework not only validates and extends perturbative and numerical findings but also offers predictive power for engineering robust edge modes in Chern insulators.

Beyond its topological utility, the dimensional‐reduction approach also casts light on hidden symmetries in long‐range‐hopping SSH chains. In Ref.~\cite{dias2022long}, it was shown that adding second‐ and third‐neighbor hoppings to the SSH model can obscure the underlying chiral and inversion symmetries unless one performs a suitable basis rotation. In our earlier plaquette‐based analysis~\cite{Rahmati}, these symmetries were not manifest, making the direct computation of the winding number less transparent. By contrast, the unitary rotation introduced in the present work aligns the dominant terms of the extended SSH Hamiltonian along Pauli axes, thereby revealing an explicit chiral (and, where present, inversion) symmetry for each $k_x$. This rotated frame not only streamlines the calculation of the winding number but also clarifies how long‐range hoppings preserve or break these key symmetries. As a result, our method provides both a topological and a symmetry‐based understanding of edge‐state protection, which was previously hidden in the unrotated formulation.

The remainder of this paper is organized as follows. In Sec.~\ref{Sec02}, we introduce the Haldane model on a honeycomb lattice, perform a dimensional reduction along the zigzag edge, and apply a unitary rotation to reveal the resulting extended SSH structure. Sec.~\ref{Sec03} delivers a detailed topological and spectral analysis: we define and compute the winding number for each $k_x$, derive analytic expressions for edge‐state wavefunctions and dispersions (with and without the mass term), identify critical momenta $k_c$, and establish the full phase diagram. Finally, Sec.~\ref{Sec04} presents a summary of our main results and discusses implications for symmetry‐protected phases in systems with long‐range hopping.


\begin{figure}[t]
\begin{center}
\includegraphics[scale=0.4] {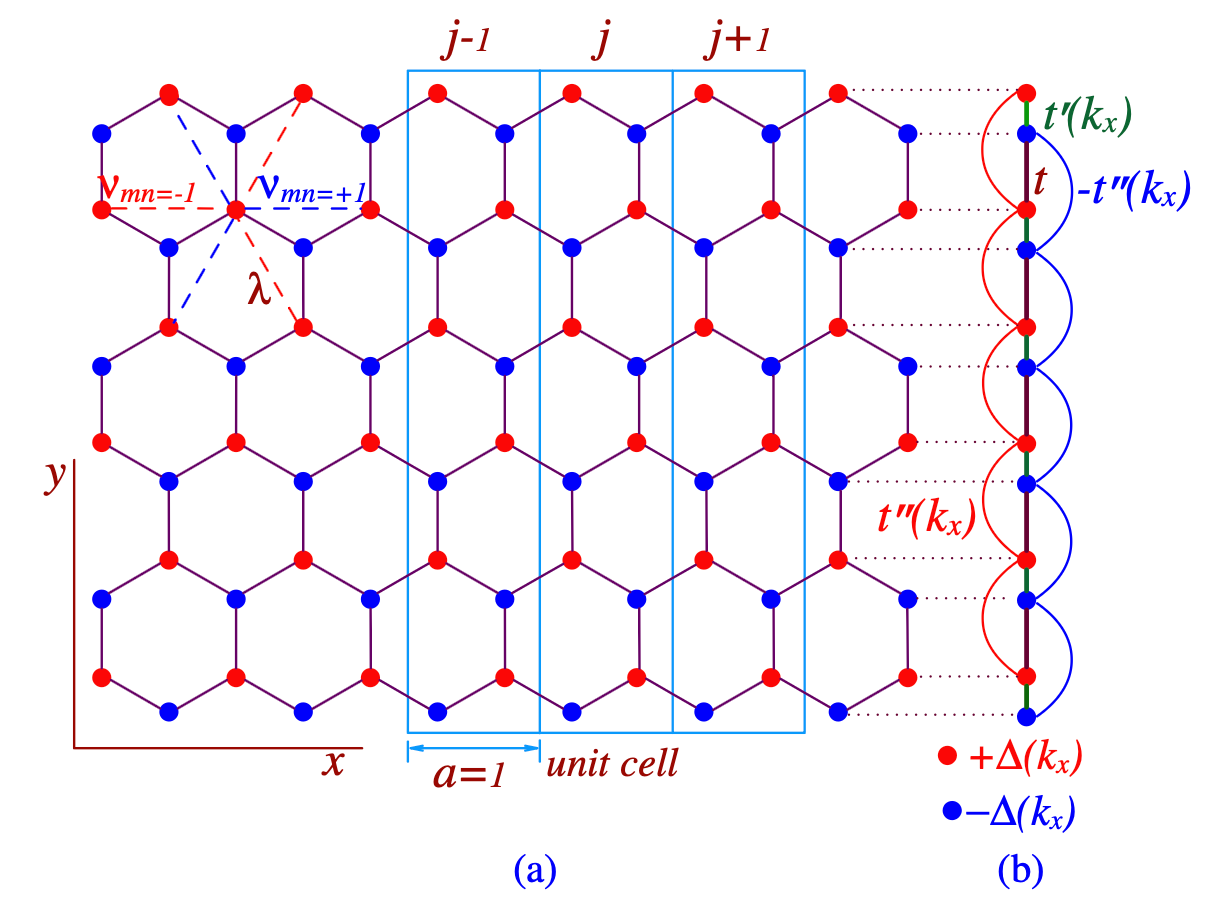}
\end{center}
\caption{(a) A schematic illustration of a 2D ribbon featuring a honeycomb lattice structure 
with zigzag edges. The ribbon extends infinitely in the $x$-direction, while it has a finite width 
in the $y$-direction. The nearest-neighbor hopping parameter is denoted as $t$, while the second-nearest 
hopping parameter $\lambda$ varies depending on the direction of hopping. The $j$-th conventional unit cell 
of the ribbon and its neighbors are indicated by rectangle boxes with a width $a = 1$.
(b) A depiction of the corresponding SSH chain derived from the ribbon geometry, parameterized by $k_x$ 
following the application of the Fourier transformation. The momentum-dependent hopping parameters 
$t'(k_x)$, $t''(k_x)$, and the on-site energies, $\pm \Delta(k_x)$, are represented using various lines 
and colors. 
\label{Fig01}}
\end{figure}
\section{One-dimensional reduction of the Haldane model~\label{Sec02}} 
The Haldane model, defined on a two-dimensional honeycomb lattice with complex next-nearest-neighbor hoppings and a staggered magnetic flux, serves as a canonical example of a Chern insulator. In this section, we consider the Haldane model implemented on a nanoribbon geometry with zigzag edges. The underlying lattice retains translational invariance along the ribbon direction ($x$-axis), while being finite in the transverse direction ($y$-axis). This setup allows for the investigation of both bulk and edge properties within a quasi-one-dimensional framework. 
We provide a detailed formulation of the tight-binding Hamiltonian and specify the relevant parameters governing the system in the following.

\begin{figure}[t]
\begin{center}
\includegraphics[scale=0.45] {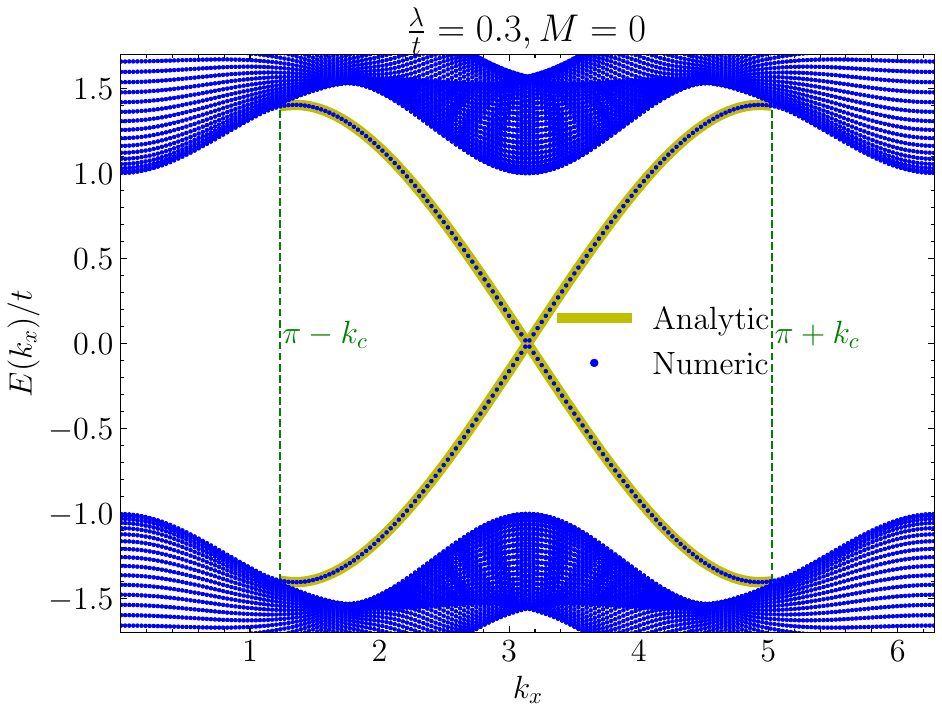}
\end{center}
\caption{
Electronic band dispersion for the chiral edge states in a zigzag ribbon of the parent Haldane model, computed numerically (blue dashed dots) for $M=0$ and $\lambda=0.3\,t$. The analytical energy dispersion of Eq.~\eqref{Eq34} obtained for the reduced SSH model (plotted as a solid yellow line) exactly matches the numerical results. Vertical dashed lines mark the points $k_x=\pi\pm k_c$, where the edge bands connect the valence and conduction bands.\label{Fig01N}}
\end{figure}

\subsection{Parent Hamiltonian Model}

A schematic representation of the system under consideration is shown in Fig.~\ref{Fig01}. 
Specifically, Fig.~\ref{Fig01}(a) illustrates a nanoribbon structure with zigzag edges, 
where the Haldane model is defined on a honeycomb lattice. 
Owing to the bipartite nature of this lattice, systems such as graphene can exhibit 
Chern insulating behavior under appropriate modifications~\cite{Haldane1988, EXP1}. 
In particular, breaking time-reversal symmetry through a staggered magnetic flux 
and lifting inversion symmetry via a sublattice-resolved mass term allows for 
nontrivial topological phases to emerge.

The tight-binding Hamiltonian that captures the essential physics of this system is given by
\begin{equation}\label{Eq01}
\begin{split}
{\cal H}= t \sum\limits_{\langle m,n\rangle}^{}  c_m^\dagger c_n +
\lambda \sum\limits_{\langle\langle m,n\rangle\rangle}^{}
e^{i\nu_{mn}\varphi} c_m^\dagger c_n\\ + M \sum_{m}^{} \xi_m c_m^\dagger c_m + \text{h.c.},
\end{split}
\end{equation}
where $t$ and $\lambda$ are the nearest-neighbor and next-nearest-neighbor hopping amplitudes, respectively. 
The notation $\langle m,n \rangle$ and $\langle\langle m,n \rangle\rangle$ refers to the respective site pairs. 
The operators $c_m^\dagger$ and $c_m$ denote fermionic creation and annihilation operators at site $m$. 
The phase $\varphi$ encodes the complex hopping induced by the magnetic flux, 
with the factor $\nu_{mn} = \pm 1$ representing the chirality of the loop formed by the next-nearest-neighbor hopping path. 
Specifically, for a triangle traversed in a clockwise (counterclockwise) manner around an A (B) sublattice site, 
$\nu_{mn}$ takes the value $+1$ ($-1$), thereby embedding the effect of the staggered magnetic field into the model. 
The third term in the Hamiltonian introduces a staggered on-site energy of magnitude $M$, 
which breaks inversion symmetry. The sublattice-dependent coefficient $\xi_m = +1$ for A sites and $-1$ for B sites 
ensures the correct sublattice asymmetry. Finally, the Hermitian conjugate term, denoted by \text{h.c.}, 
ensures that the full Hamiltonian remains Hermitian.


The Haldane model exhibits a topological phase transition driven by the competition between two key effects. The complex next-nearest-neighbor hopping, which breaks time-reversal symmetry, induces effective mass terms at the Dirac points with opposite signs, favoring a topologically nontrivial phase with a nonzero Chern number $\mathcal{C}\neq0$. In contrast, the inversion symmetry–breaking (Semenoff) mass term provides a uniform mass for both Dirac points, tending to drive the system into a trivial insulating phase. A phase transition occurs when the gap closes—at one of the Dirac points—and subsequently reopens with an inversion of the mass sign at that point. In a certain parameter regime, the two Dirac points have mass terms of opposite signs, yielding a nontrivial Chern insulating phase with $\mathcal{C}=1$. When the staggered potential exceeds a critical strength, both Dirac points acquire masses of the same sign, and the system becomes a trivial insulator with $\mathcal{C}=0$. This change is reflected in the appearance or disappearance of chiral edge states, consistent with the bulk-boundary correspondence. For simplicity, in what follows we consider the special case $\varphi = \pi/2$, for which the phase boundary is determined by $M = \pm 3\sqrt{3}\lambda$.

\subsection{Reduced Hamiltonian: Extended SSH model}
To proceed with our analysis, we take advantage of the translational invariance of the system along the $x$-axis, as shown in Fig.~\ref{Fig01}(a). This symmetry implies that the Bloch wave number $k_x$ is a good quantum number and allows for a partial Fourier transform along the $x$-direction. As a result, for each fixed value of $k_x$, the original two-dimensional Hamiltonian in Eq.~\ref{Eq01} can be mapped onto an effective one-dimensional chain extending along the $y$-axis~\cite{Rahmati, Sadeghizadeh, Sodagar, Tan}, as schematically depicted in Fig.~\ref{Fig01}(b). 

This effective one-dimensional system can be interpreted as a generalized Su-Schrieffer-Heeger (SSH) chain, where each conventional unit cell of the Haldane model is projected into an extended unit cell aligned along the $y$-direction. The corresponding $k_x$-dependent Hamiltonian can be expressed in the compact form:
\begin{equation}\label{Eq02}
\begin{split}
{\cal H}(k_x)= \sum_{i} &\Big(\begin{array}{cc}
                                c_{iA}^\dagger &   c_{iB}^\dagger
                             \end{array}\Big)
\Big[h_{i-1,i}(k_x)\\ &+h_{i,i}(k_x)+h_{i,i+1}(k_x)\Big]\Big(\begin{array}{c}
                                c_{iA}\\   c_{iB}
                             \end{array}\Big)+\text{h.c.},
\end{split}
\end{equation}
where $i$ indexes the unit cells along the $y$-direction, and $c_{iA}^\dagger$, $c_{iB}^\dagger$, $c_{iA}$, and $c_{iB}$ are the electron creation and annihilation operators for sublattices A and B within the $i$-th unit cell. The terms $h_{i,i}(k_x)$, $h_{i-1,i}(k_x)$, and $h_{i,i+1}(k_x)$ correspond to the intracell and intercell coupling matrices, respectively. The intracell component $h_{i,i}(k_x)$ accounts for all interactions confined within a single unit cell, while $h_{i\pm1,i}(k_x)$ describe hopping processes between neighboring cells. These components are explicitly given by:
\begin{equation}\label{Eq03}
\begin{split}
h_{i,i}(k_x)  & = t'(k_x)  \sigma_x +\big[ M + \Delta (k_x)\big] \sigma_z, \\
h_{i,i+1} (k_x)   & = t'' (k_x) \sigma_z + t\,\sigma^+,\\
h_{i-1,i} (k_x)   & = h_{i,i+1}^\dagger(k_x),
\end{split}
\end{equation}
where $\sigma_x$ and $\sigma_z$ denote the standard Pauli matrices, and $\sigma^+$ is the Pauli raising operator. The momentum-dependent hopping parameters and staggered potential contributions are given by:
\begin{equation}\label{Eq04}
\begin{split}
t'(k_x) & = 2 t \cos(k_x/2),\\  
t'' (k_x) & = 2 \lambda \sin(k_x/2), \\ 
\Delta(k_x) & = 2 \lambda \sin(k_x).
\end{split}
\end{equation}
Throughout this work, we adopt the convention that the lattice spacing is set to unity ($a=1$), unless otherwise specified.

\begin{figure*}[t]
\begin{center}
\includegraphics[scale=0.6] {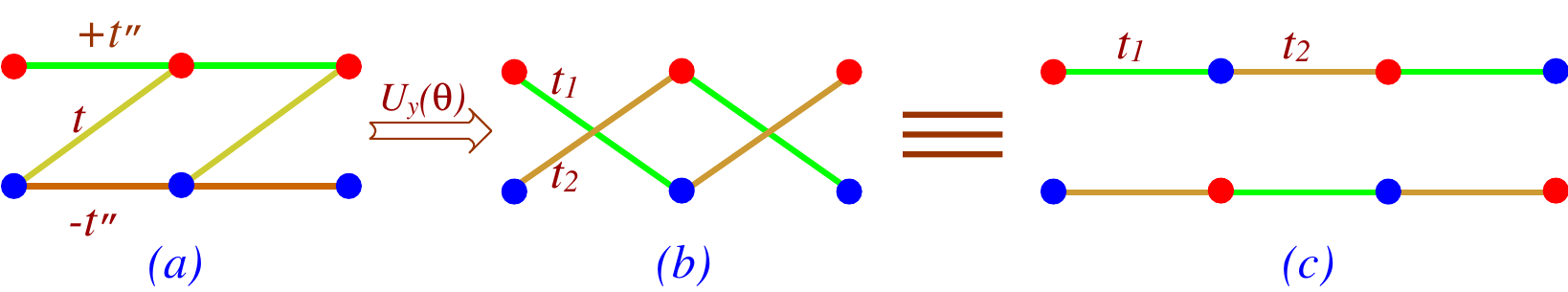}
\end{center}
\caption{(a) At the point of $k_x=\pi$ and in the absence of the mass term in the Hamiltonian, 
the reduced system corresponds to an extended SSH chain depicted in the figure, in this representation the hopping 
parameter $t''$ is equal to $2\lambda$. (b) By applying a unitary rotation 
around the $y$-axis by an angle of $\theta$, the Hamiltonian transforms into a form that corresponds 
to two independent 1D~SSH structures. (c) By rearranging the atoms, these two chains are perceived as 
two completely independent chains, one of which possesses an edge state while the other does not.
In fact, (c) is equivalent to (b), only the atoms have been rearranged to show the independence 
of the two chains more clearly. \label{Fig02}}
\end{figure*}
\subsection{Topological Characterization of the Extended SSH Model}

In order to explore the topological features of the extended SSH chain obtained in the previous section, we now analyze its bulk properties in momentum space. Although the physical system derived from the nanoribbon geometry has finite extent along the $y$-direction, we momentarily assume it to be infinitely extended. This allows us to focus on the bulk characteristics, following the standard approach used in the study of the original SSH model.

For each fixed value of $k_x$, the effective one-dimensional chain aligned along the $y$-axis can be transformed to momentum space by applying a Fourier transform along the $y$-direction. This yields a momentum-resolved Hamiltonian $H(k_y)$, which takes the following form:
\begin{equation}\label{Eq05}
\begin{split}
H(k_y)= [t \cos k_y + t'(k_x)] \sigma_x + [t \sin k_y]\sigma_y \\
\quad + [M+\Delta(k_x) + 2 t''(k_x) \cos k_y]\,\sigma_z.
\end{split}
\end{equation}

This Hamiltonian is naturally expressed in terms of Pauli matrices, making it amenable to a geometric interpretation in the Bloch sphere framework. In particular, it can be compactly written as:
\begin{equation}\label{Eq06}
H(k_y) = \mathbf{d}(k_y)\cdot\boldsymbol{\sigma}
= d_x(k_y) \sigma_x + d_y(k_y) \sigma_y + d_z(k_y) \sigma_z,
\end{equation}
where the vector $\mathbf{d}(k_y) = (d_x, d_y, d_z)$ characterizes the system's band geometry and encodes its topological nature. Here, the components of $\mathbf{d}(k_y)$ are given by:
\begin{equation}
\begin{split}
d_x(k_y) &= t \cos k_y + t'(k_x),\\
d_y(k_y) &= t \sin k_y,\\
d_z(k_y) &= M + \Delta(k_x) + 2 t''(k_x) \cos k_y.
\end{split}
\end{equation}

To characterize the topological nature of the bulk bands, it is crucial to treat the Hamiltonian as a continuous mapping from the 1D Brillouin zone, parametrized by $k_y \in [-\pi, \pi]$, to the Bloch sphere. For this mapping to be well-defined, the normalized vector $\hat{\mathbf{d}}(k_y) = \mathbf{d}(k_y)/|\mathbf{d}(k_y)|$ must trace a closed loop on the surface of the Bloch sphere as $k_y$ spans the Brillouin zone. 
The winding number ($\nu$), as our topological invariant, quantifies the number of times this loop winds around the origin and serves as a distinguishing marker between topologically trivial and nontrivial phases. It can be defined only in the presence of chiral symmetry~\cite{RMP4} and
mathematically, reads as~\cite{short}:
\begin{equation}\label{Eq07}
 \nu = \frac{1}{2\pi} \int_{-\pi}^{\pi} 
\left[\hat{\mathbf{d}}(k_y) \times \frac{d\hat{\mathbf{d}}(k_y)}{dk_y}\right] \cdot \hat{\mathbf{n}}\, dk_y,
\end{equation}
where $\hat{\mathbf{n}}$ is a fixed unit vector pointing along the direction of the mapping on the Bloch sphere. The requirement that the Hamiltonian be expressed in terms of Pauli matrices is essential for this construction, as it makes explicit the geometric structure of the band eigenstates and enables the computation of the $\nu$. Ultimately, this invariant captures the topological information encoded in the bulk band structure and governs the emergence of edge states through the principle of bulk-boundary correspondence.


Before proceeding to the details of the $\nu$ analysis, we note that, by inspection of Eq.~\ref{Eq05}, the Bloch‐sphere mapping
is in fact confined to a specific two‐dimensional plane within the three‐dimensional $d$‐space.  Indeed, although all three components 
$d_x(k_y),\; d_y(k_y),\; d_z(k_y)$
are generally nonzero, they satisfy a linear relation that forces their trajectory to lie in a tilted plane.  This geometric constraint ensures that, as $k_y$ runs from $-\pi$ to $+\pi$, the vector $\hat{\mathbf{d}}(k_y)$ traces a closed loop on the Bloch sphere which is guaranteed to be planar.

To make use of this property and simplify the winding‐number computation, we perform a basis rotation that brings the entire loop into a plane parallel to the $xy$–plane.  Concretely, one introduces the rotated Hamiltonian
\begin{equation}\label{Eq08}
\tilde{H}(k_y)
= \tilde{\mathbf{d}}(k_y)\cdot\boldsymbol{\sigma}
= \tilde{d}_x(k_y)\,\sigma_x + \tilde{d}_y(k_y)\,\sigma_y + \tilde{d}_z\,\sigma_z,
\end{equation}
where by construction $\tilde{d}_z$ is independent of $k_y$ (or zero), so that $\tilde{\mathbf{d}}(k_y)$ executes a planar loop in the $xy$–plane.  
This form allows the evaluation of the winding number immediately. 

To achieve Eq.~\ref{Eq08}, we rotate the dominant terms in the original Hamiltonian of Eq.~\ref{Eq05} about the $y$–axis.  Writing
$t\,\sigma_x + 2\,t''(k_x)\,\sigma_z$
and choosing the rotation angle
\begin{equation}\label{Eq09}
\theta = \tan^{-1}\!\Bigl(\frac{2\,t''(k_x)}{t}\Bigr)
= \tan^{-1}\!\Bigl(\frac{4\lambda\sin(k_x/2)}{t}\Bigr),
\end{equation}
one finds that
\begin{equation}\label{Eq10}
\tilde{t}(k_x)
= \sqrt{\,t^2 + \bigl[2\,t''(k_x)\bigr]^2\,}
= \sqrt{t^2 + 16\,\lambda^2\,\sin^2\!(k_x/2)}
\end{equation}
is the resulting amplitude along the rotated $\sigma_x$ direction.  This rotation is implemented by
\begin{equation}\label{Eq12}
U_y(\theta) = e^{-i\sigma_y\,\theta/2}
= \begin{pmatrix}
\cos(\theta/2) & -\sin(\theta/2) \\
\sin(\theta/2) &  \cos(\theta/2)
\end{pmatrix},
\end{equation}
which ensures
\begin{equation}\label{Eq11}
U_y(\theta)\bigl[\,2t''(k_x)\,\sigma_z + t\,\sigma_x\bigr]U_y^\dagger(\theta)
= \tilde{t}(k_x)\,\sigma_x.
\end{equation}

Applying the same unitary $U_y(\theta)$ to the full Hamiltonian in Eq.~\ref{Eq05} yields the rotated components
\begin{equation}\label{Eq13}
\begin{split}
\tilde{d}_y(k_y) &= d_y(k_y) = t\sin k_y,\\
\tilde{d}_x(k_y) &= -M_x(k_x) + \tilde{t}(k_x)\,\cos k_y,\\
\tilde{d}_z     &= M_z(k_x),
\end{split}
\end{equation}
where the $k_x$–dependent shifts are
\begin{equation}\label{Eq14}
\begin{split}
M_x(k_x) &= \bigl[M + \Delta(k_x)\bigr]\sin\theta \;-\; t'(k_x)\cos\theta,\\
M_z(k_x) &= \bigl[M + \Delta(k_x)\bigr]\cos\theta \;+\; t'(k_x)\sin\theta.
\end{split}
\end{equation}
By construction, $\tilde{\mathbf{d}}(k_y)$ now lies strictly in a horizontal plane, so that the winding number can be read off directly as the number of times the closed loop encircles the origin in the $xy$–plane. This completes the transformation that makes the topological analysis fully transparent and closely analogous to the original SSH model.


\begin{figure}[t]
\begin{center}
\includegraphics[scale=0.45] {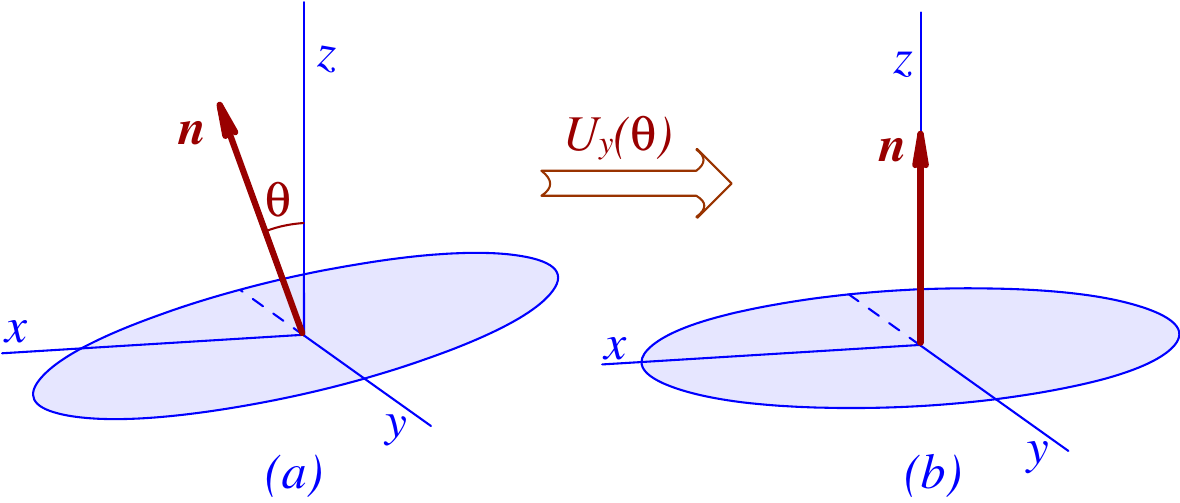}
\end{center}
\caption{(a) In the absence of a the mass term, vector $\mathbf{d}$ covers a loop in a plane that passes 
through the origin and is perpendicular to the vector $\mathbf{n}$. The plane in which the loop is placed 
includes the $y$-axis and the closed path is symmetrical with respect to this axis. (b) To calculate the 
$\nu$, it is easier to align the vector $\mathbf{n}$ along the $z$-axis using a unitary rotation.  
This alignment results in the vector $\mathbf{d}$ sweeping a loop in the $xy$-plane thereby simplifying 
the calculation of the $\nu$. \label{Fig03}}
\end{figure}



Before ending this section, an important question naturally arises: which real‐space tight‐binding model reproduces the rotated Hamiltonian $\tilde{H}(k_y)$?  One finds that it emerges from a one‐dimensional chain with the following site‐ and bond‐dependent couplings:
\begin{equation}\label{Eq15}
\begin{split}
\tilde{h}_{i,i}(k_x)    &= -M_x(k_x)\,\sigma_x \;+\; M_z(k_x)\,\sigma_z, \\
\tilde{h}_{i,i+1}(k_x)  &= t_1(k_x)\,\sigma^+ \;+\; t_2(k_x)\,\sigma^-, \\
\tilde{h}_{i-1,i}(k_x)  &= \tilde{h}_{i,i+1}^\dagger(k_x),
\end{split}
\end{equation}
where $\sigma^\pm = (\sigma_x \pm i\,\sigma_y)/2$ and the effective hopping amplitudes are
\begin{equation}\label{Eq16}
\begin{split}
t_1(k_x) &= \frac{\tilde{t}(k_x) - t}{2}, \\
t_2(k_x) &= \frac{\tilde{t}(k_x) + t}{2}.
\end{split}
\end{equation}
In this representation, $M_x(k_x)$ governs the onset of topological phase transitions by controlling the gap‐closing condition, while $M_z(k_x)$ sets the energy of any localized edge mode that may arise.

To summarize the key takeaways from this rotated‐Hamiltonian analysis:  
1. The winding number ($\nu$) is rigorously defined and directly computable via the planar loop of $\tilde{\mathbf{d}}(k_y)$.  
2. A nonzero $\nu$ coincides with the nontrivial Chern insulating phase of the original 2D model; when the system crosses its phase boundary, the $\nu$ drops to zero, mirroring the change in the Chern number.  
3. The effective 1D model admits exact analytical expressions for both the bulk dispersion and the edge‐state wave functions.  
4. In the full nanoribbon spectrum (see Fig.~\ref{Fig01N}), the edge bands intersect the bulk at a well‐defined momentum $k_c$; our framework yields a closed‐form expression for this critical $k_c$.  

These results demonstrate both the analytical transparency and the topological depth of the dimensionally reduced SSH‐like model, paving the way for the detailed phase‐diagram and edge‐state studies in the sections that follow.

\begin{figure*}[t]
\begin{center}
\includegraphics[scale=0.6] {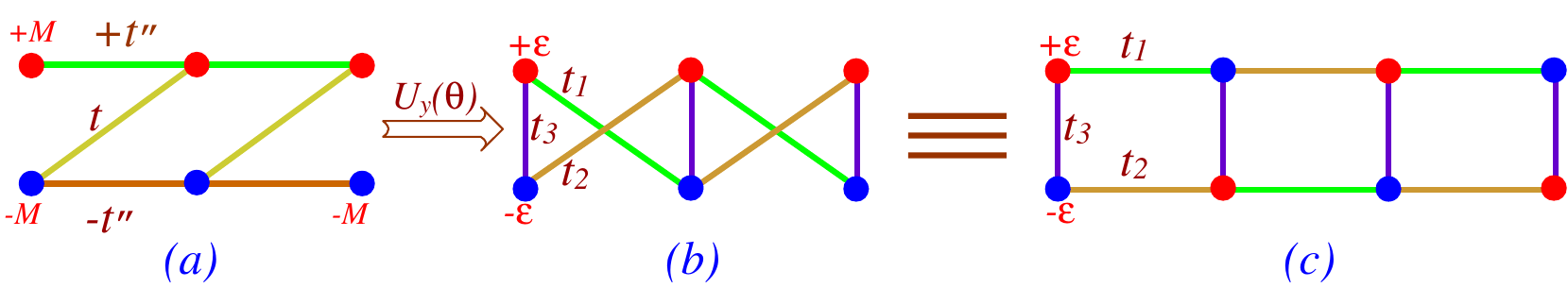}
\end{center}
\caption{(a) Same as Fig.~\ref{Fig02}, but with the inclusion of the mass term in the Hamiltonian. 
The presence of the mass term induces the energies of $+M$ and $-M$ on the sites of sublattices A 
and B, respectively. (b) By rotating the Hamiltonian around 
the $y$-axis, two SSH chains are produced, introducing $t_3=-M \sin\theta$ as the hopping parameter between the adjacent 
atoms of the chains and differing on-site energies $\pm \epsilon=\pm M\cos\theta$ at the A and B sites. (c) Rearranging the atoms allows 
for a clearer display of the structure of the SSH chains. \label{Fig04}}
\end{figure*}


\section{Results and discussion~\label{Sec03}}

In this section, we present the analytical findings that arise from the framework developed above.  To organize the discussion clearly, we divide our results into two parts.  In Section~\ref{Sec03a}, we focus on the special point $k_x = \pi$, where Eq.~\ref{Eq04} yields the simplifications $t'(\pi) = 0$, $\Delta(\pi) = 0$, and $t''(\pi) = 2\lambda$, leading to a highly tractable extended SSH model.  In Section~\ref{Sec03b}, we then generalize to arbitrary values of $k_x$, examining how the winding number, bulk gap, and edge‐state dispersion evolve throughout the Brillouin zone.  This two‐fold approach highlights both the illustrative simplicity at $k_x = \pi$ and the richer behavior encountered away from this high‐symmetry point.

\subsection{Case Study at $k_x = \pi$~\label{Sec03a}}

\subsubsection{Topological properties for $M=0$}
As we already discussed, at $k_x = \pi$ the relations in Eq.~\ref{Eq04} simplify the effective extended SSH Hamiltonian introduced in Eq.~\ref{Eq03}.  The Hamiltonian components at $k_x = \pi$ reduce to:
\begin{equation}\label{Eq17}
\begin{split}
h_{i,i}(k_x=\pi)    & = M \,\sigma_x, \\
h_{i,i+1} (k_x=\pi) & = 2\lambda \,\sigma_z \;+\; t \,\sigma^+,\\
h_{i-1,i} (k_x=\pi) & = h_{i,i+1}^\dagger (k_x=\pi).
\end{split}
\end{equation}
To isolate the topological features, we first set the mass term $M$ to zero.  In this limit the onsite term vanishes, $h_{i,i}(k_x = \pi) = 0$, and the system is known to carry a Chern number of unity, $\mathcal{C}=1$.  This topologically nontrivial phase provides a clear benchmark for analyzing both the edge‐state structure and the winding behavior.  A schematic of the resulting extended SSH lattice is shown in Fig.~\ref{Fig02}(a).  This concrete example at $k_x = \pi$ will guide our subsequent extension to general $k_x$ values.

Applying a Fourier transform along the $y$–direction to the simplified Hamiltonian at $k_x = \pi$ yields the momentum–space form
\begin{equation}\label{Eq18}
H(k_y) = (t \cos k_y)\,\sigma_x + (t \sin k_y)\,\sigma_y + (4\lambda \cos k_y)\,\sigma_z,
\end{equation}
in which all three components of $\mathbf{d}(k_y)$ are nonzero.  As depicted in Fig.~\ref{Fig03}(a), the loop traced by this vector is symmetric about the $y$–axis but lies in a plane tilted around that axis, intersecting the $xy$–plane only along the $y$–direction.  To restore planarity, we perform a unitary rotation about the $y$–axis by the angle
\[
\theta = \tan^{-1}\!\bigl(\tfrac{4\lambda}{t}\bigr),
\]
after which the Hamiltonian becomes
\begin{equation}\label{Eq19}
\tilde{H}(k_y) = \bigl[\tilde{t}\cos k_y\bigr]\,\sigma_x + \bigl[t\sin k_y\bigr]\,\sigma_y,
\end{equation}
with 
\[
\tilde{t} = \sqrt{16\lambda^2 + t^2}.
\]
In this rotated frame, the vector $\tilde{\mathbf{d}}(k_y)$ describes an ellipse centered at the origin in the $xy$–plane (Fig.~\ref{Fig03}(b)), having semi–major axis $\tilde{t}$ along $x$ and semi–minor axis $t$ along $y$.  Because this ellipse winds once around the origin as $k_y$ traverses $[-\pi,\pi]$, the winding number is unambiguously $\nu=1$.

One can further show that $\tilde{H}(k_y)$ derives from a real–space tight–binding chain with components
\begin{equation}\label{Eq20}
\begin{split}
\tilde{h}_{i,i}(k_x=\pi)    &= 0,\\
\tilde{h}_{i,i+1}(k_x=\pi)  &= t_1\,\sigma^+ + t_2\,\sigma^-,\\
\tilde{h}_{i-1,i}(k_x=\pi)&= \tilde{h}_{i,i+1}^\dagger(k_x=\pi),
\end{split}
\end{equation}
where
\begin{equation}\label{Eq21}
t_1 = \frac{\tilde{t} - t}{2}, \qquad
t_2 = \frac{\tilde{t} + t}{2}.
\end{equation}
As illustrated in Fig.~\ref{Fig02}(b), the unitary rotation decouples the extended SSH chain into two independent 1D SSH chains.  Upon rearranging the sites (Fig.~\ref{Fig02}(c)), each chain displays alternating hoppings $t_1$ and $t_2$.  Since $t_1 < t_2$, only the lower chain hosts topological edge modes, while the upper chain remains trivial.

\subsubsection{Edge state analysis for $M=0$}

In the $M=0$ limit, only the lower SSH chain supports a localized edge mode whose wave function can be written in the on‐site basis $\{\lvert n,A,1\rangle\}$ as
\begin{equation}\label{Eq22}
\lvert\psi_{\rm edge}\rangle
= \frac{1}{c}\sqrt{1 - c^2}\sum_{n=1}^\infty c^n\,\lvert n,A,1\rangle,
\end{equation}
with decay constant
\begin{equation}\label{Eq23}
c = -\frac{t_1}{t_2} = -\frac{\tilde{t}-t}{\tilde{t}+t}\,.
\end{equation}
Here the amplitude at $n=1$ is unity and falls off as $c^{\,n-1}$.  Transforming back to the original two‐component basis of the extended chain introduces a fixed rotation by $\theta = \tan^{-1}(4\lambda/t)$, yielding
\begin{equation}\label{Eq24}
\lvert\psi_{\rm edge}\rangle
= \frac{1}{c}\sqrt{1 - c^2}\sum_{n=1}^\infty c^n
\begin{bmatrix}
\cos(\theta/2)\,\lvert n,A,1\rangle\\
\sin(\theta/2)\,\lvert n,B,2\rangle
\end{bmatrix}.
\end{equation}
This form makes explicit the exponential localization of the edge state and its sublattice‐spinor structure in the original Haldane‐ribbon basis.

\begin{figure}[t]
\begin{center}
\includegraphics[scale=0.45] {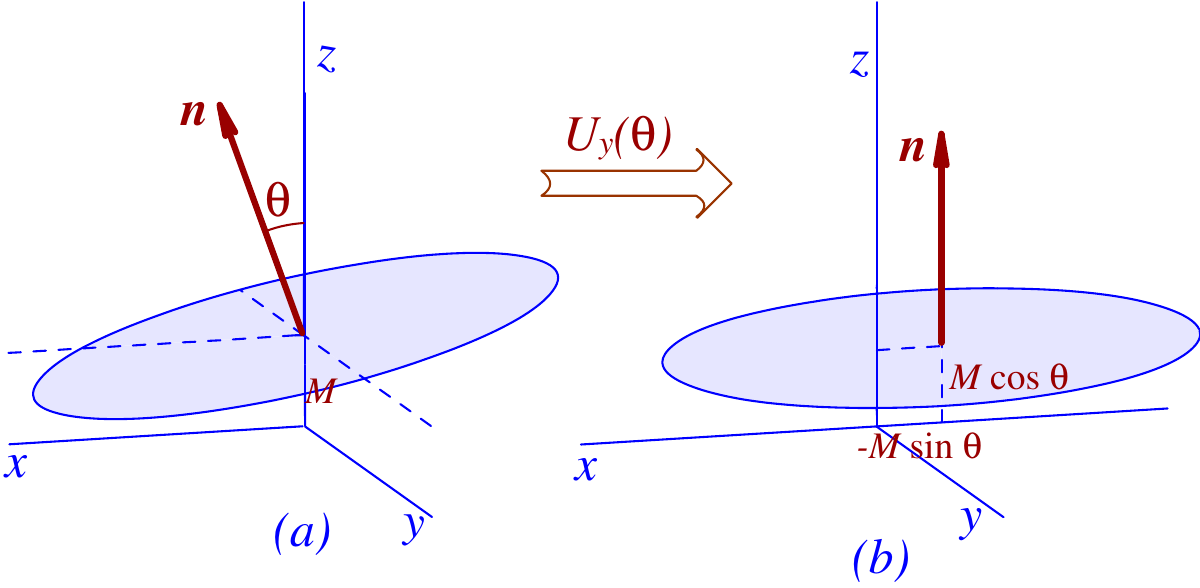}
\end{center}
\caption{(a) Sams as Fig.~\ref{Fig02}, but with the inclusion of the mass term in the Hamiltonian. 
The $k$-space Hamiltonian is situated in a plane that is rotated 
relative to the $xy$-plane. Additionally, the center of the closed loop traced by the vector $\mathbf{d}$
is located at $z=+M$. (b) By performing a unitary rotation of the Hamiltonian around the $y$-axis 
by an angle $\theta$, it can be mapped onto o horizental plane. However, in this case, the center of 
the loop traced by the vector $\tilde{\mathbf d}$ does not coincide with the origin of the coordinates, instead, 
it is shifted along the $x$-axis by $-M\sin\theta$, while its coordinate has decreased to $M\cos\theta$ in 
the direction of $z$. \label{Fig05}}
\end{figure}
\subsubsection{Topological properties for non-zero mass}

Introducing a finite mass term at $k_x = \pi$ alters both the Bloch–space Hamiltonian and its real–space tight–binding representation.  In momentum space one now has
\begin{equation}\label{Eq25}
\begin{split}
\tilde{H}(k_y) 
&= \bigl[-M \sin\theta + \tilde{t} \cos k_y\bigr]\sigma_x 
+ t \sin k_y\,\sigma_y \\
&\quad + M \cos\theta\,\sigma_z,
\end{split}
\end{equation}
where the constant $M\cos\theta$ term shifts the loop center vertically, and the ellipse traced by $\tilde{\mathbf{d}}(k_y)$ lies in a horizontal plane but is displaced to $(x,y,z)=(-M\sin\theta,0,M\cos\theta)$ with axes $\tilde{t}$ (along $x$) and $t$ (along $y$) (see Fig.~\ref{Fig05}(b)).  

In the corresponding real–space chain, the rotated Hamiltonian’s components become
\begin{equation}\label{Eq26}
\begin{split}
\tilde{h}_{i,i}(k_x=\pi)   
&= -M\sin\theta\,\sigma_x + M\cos\theta\,\sigma_z,\\
\tilde{h}_{i,i+1}(k_x=\pi) 
&= t_1\,\sigma^+ + t_2\,\sigma^-,\\
\tilde{h}_{i-1,i}(k_x=\pi) 
&= \tilde{h}_{i,i+1}^\dagger(k_x=\pi),
\end{split}
\end{equation}
where the extra $\sigma_x$ on–site term encodes the horizontal shift and the $\sigma_z$ term the vertical one.

Equivalently, in the language of two coupled SSH chains (Fig.~\ref{Fig04}(b)–(c)), the mass term introduces:
1. An interchain hopping $t_3 = -M\sin\theta$ connecting A– and B–sites across the chains, and  
2. A staggered on–site potential $\pm\epsilon$ with $\epsilon = M\cos\theta$ on sublattices A/B.

Despite these modifications, the loop in Fig.~\ref{Fig05} still encloses the origin for sufficiently small $M$, implying a nonzero $\nu$ and the persistence of an edge mode (with vanishing amplitude on B sublattice sites).  The full tight–binding Hamiltonian $H(k_y)$ thus contains alternating on–site energies $\pm\epsilon$, intra–chain hoppings $t_1,t_2$, and interchain coupling $t_3$, and supports a localized edge state until the mass reaches the critical value that drives the winding number to zero.

\begin{figure*}[t]
\begin{center}
\includegraphics[scale=0.6] {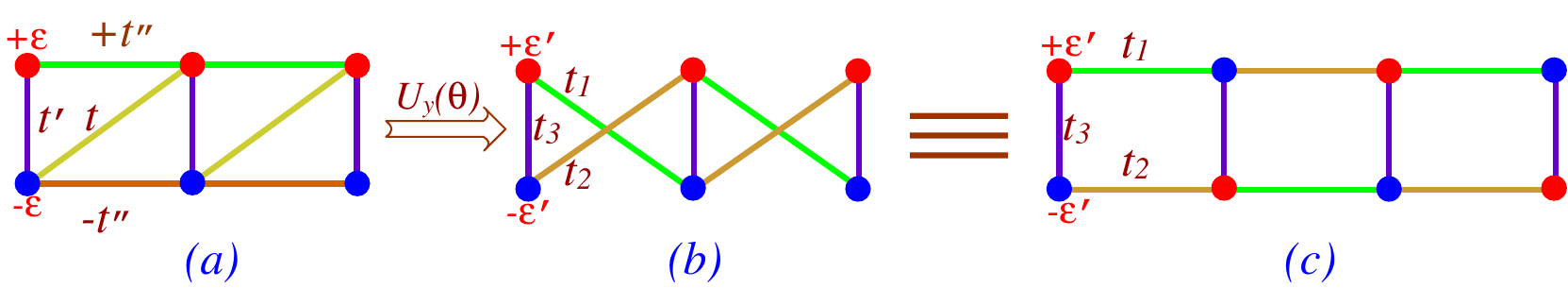}
\end{center}
\caption{(a)~Same as Fig.~\ref{Fig04}, but for any arbitrary value of  $k_x$ in the first Brillouin zone. 
The presence of the mass term and non-zero $\Delta$ induces $\pm\epsilon=\pm M \pm \Delta$
energies on the A and B sub-lattice sites, respectively. Additionally, the hopping parameter $t'$
is non-zero in this case. (b)~Same as Fig.~\ref{Fig04}(b), but 
$t_3= t' \cos\theta-[M+\Delta]\sin\theta$ and $\epsilon'=-t'\sin\theta+[M+\Delta]\cos\theta$.
(c)~Rearranging the atoms allows for a clearer display of the structure of the 1D SSH chains.
\label{Fig06}}
\end{figure*}

\subsubsection{Edge State Analysis for non-zero mass}

We now extend the edge‐state construction to the case of a finite mass term at $k_x = \pi$.  Assuming a localized mode exists, its wave function can be written as a superposition over A–sublattice sites of both SSH chains:
\begin{equation}\label{Eq27}
\lvert\psi_{\rm edge}\rangle
= N \sum_{n=1}^{\infty}\Bigl(\alpha_n\,\lvert n,A,1\rangle + \beta_n\,\lvert n,A,2\rangle\Bigr),
\end{equation}
with normalization
\begin{equation}\label{Eq28}
N = \Biggl[\sum_{n=1}^\infty\bigl(|\alpha_n|^2 + |\beta_n|^2\bigr)\Biggr]^{-1/2}.
\end{equation}
Closed‐form expressions for the amplitudes $\alpha_n$, $\beta_n$, and $N$ are derived in Appendix.~\ref{AppA}.

Rotating back into the original Haldane–ribbon basis by the same angle $\theta = \tan^{-1}(4\lambda/t)$ yields
\begin{widetext}
\begin{equation}\label{Eq29}
\lvert\psi_{\rm edge}\rangle
= N \sum_{n=1}^{\infty} \Bigg[\,
\alpha_n
\begin{pmatrix}
\cos(\tfrac\theta2)\,\lvert n,A,1\rangle\\[6pt]
\sin(\tfrac\theta2)\,\lvert n,B,2\rangle
\end{pmatrix}
\;+\;
\beta_n
\begin{pmatrix}
\sin(\tfrac\theta2)\,\lvert n,B,1\rangle\\[6pt]
\cos(\tfrac\theta2)\,\lvert n,A,2\rangle
\end{pmatrix}\!\Bigg].
\end{equation}
\end{widetext}

A key convergence criterion emerges from the recurrence relations: the edge‐state remains normalizable only if
\[
M\,\sin\theta < \tilde{t} = t_1 + t_2.
\]
When this inequality holds, $\lvert\psi_{\rm edge}\rangle$ decays exponentially and defines a stable, topologically protected mode.  As $M\sin\theta$ approaches $\tilde{t}$, one root of the recurrence hits $-1$, causing the wave function to diverge.  This signals the topological phase transition at
\[
M \sin\theta = \tilde{t}\,: 
\quad \nu:1\;\longrightarrow\;0,
\]
corresponding to the bulk gap closing (Fig.~\ref{Fig05}(b)) and the disappearance of edge localization for $M \sin\theta > \tilde{t}$.

Finally, within the nonzero–mass edge‐state subspace the term $M\cos\theta\,\sigma_z$ in the rotated Hamiltonian acts as a constant on‐site energy.  Consequently, $\lvert\psi_{\rm edge}\rangle$ is an eigenstate of $M\cos\theta\,\sigma_z$ with eigenvalue
\[
E(k_x = \pi) \;=\; M\,\cos\theta.
\]

\begin{figure}[t]
\begin{center}
\includegraphics[scale=0.4] {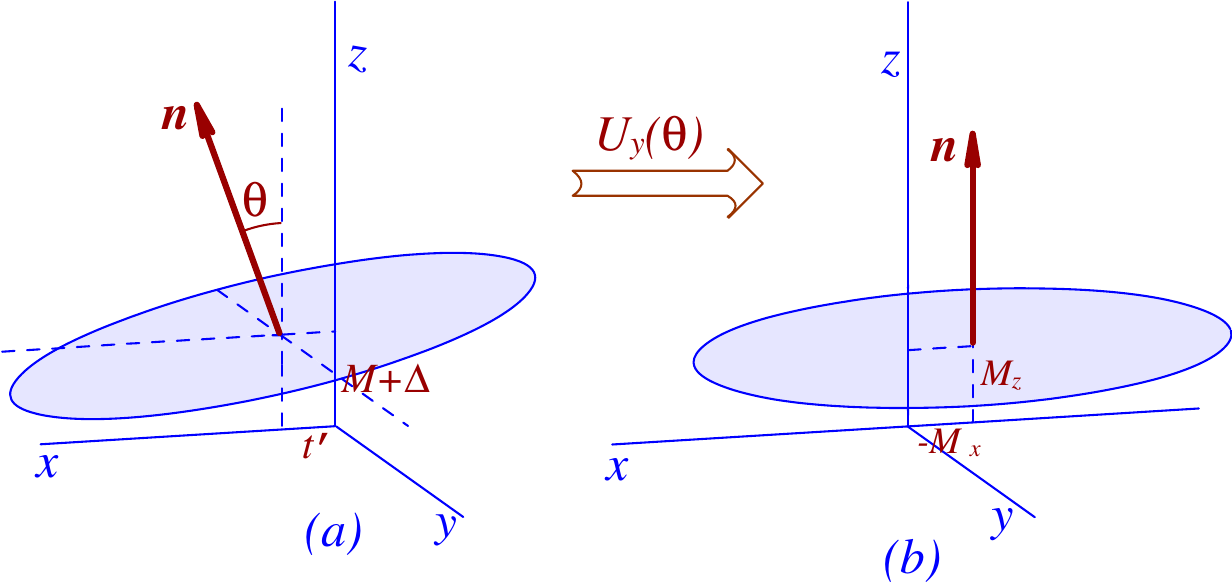}
\end{center}
\caption{(a) Sams as Fig.~\ref{Fig05}, but with the inclusion of the mass term in the Hamiltonian. 
The $k$-space Hamiltonian is situated in a plane that is rotated 
relative to the $xy$-plane. Additionally, The center of the loop traced by $\mathbf{d}$ is shifted relative to the origin by $t'$ and $M+D$ 
along the $x$ and $z$ directions, but the horizontal symmetry axis of the loop is parallel to the $y$ axis. 
(b) By performing a unitary rotation of the Hamiltonian around the $y$-axis 
by an angle $\theta$, it can be mapped onto a horizental. However, in this case, the center of 
the loop traced by the vector $\tilde{\mathbf d}$ does not coincide with the origin of the coordinates, instead, 
it is shifted along the $x$ and $z$ axes by $-M_x$ and $M_z$, respectively. \label{Fig07}}
\end{figure}
\subsection{Generic $k_x$ Analysis Beyond $\pi$~\label{Sec03b}}

We now extend our study to arbitrary values of $k_x$, using the full momentum–dependent Hamiltonian derived in Section~\ref{Sec02}.  First, for each fixed $k_x$ we compute the winding number ($\nu$) of the reduced 1D Hamiltonian by evaluating the loop of $\tilde{\mathbf{d}}(k_y)$ in its rotated frame.  Simultaneously, we construct the corresponding edge‐state wave functions and obtain their dispersion relations.  Next, we identify the critical momentum $k_c$ at which these edge bands merge with the bulk spectrum (see Fig.~\ref{Fig01N}) and derive its exact value.  Finally, we investigate how the inclusion of a finite mass term $M$ modifies both the edge‐state structure and the $\nu$, thereby yielding the analytic condition for the topological phase transition in the original Haldane nanoribbon.  This comprehensive analysis reveals the full topological phase diagram as a function of $(k_x,M)$.

In the general case $k_x\in[-\pi,\pi]$, the reduced 1D Hamiltonian retains the form given in Eq.~\ref{Eq03}, with its hopping parameters now explicit functions of $k_x$ as in Eq.~\ref{Eq04}.  The corresponding extended SSH chain is sketched in Fig.~\ref{Fig06}(a), where the sublattice on‐site energies become
\begin{equation}\label{Eq30}
\epsilon(k_x) = M + \Delta(k_x) = M + 2\lambda\sin k_x,
\end{equation}
and an additional nearest‐neighbor coupling
\[
t'(k_x) = 2t \cos\!\bigl(\tfrac{k_x}{2}\bigr)
\]
connects A and B sites across the parallel chains (absent at $k_x=\pi$).  The $k$–space Hamiltonian again takes the form of Eq.~\ref{Eq05}, and its Bloch‐vector $\mathbf{d}(k_y)$ traces a closed loop in a plane tilted relative to the $xy$–plane, centered at
\[
\bigl[t'(k_x),\,0,\,M+\Delta(k_x)\bigr],
\]
as shown in Fig.~\ref{Fig07}(a).  By performing the same unitary rotation about the $y$–axis by angle $\theta$ of Eq.~\ref{Eq09}, one maps this loop into a horizontal plane, rendering the winding number straightforward to compute.  A nonzero $\nu$ again guarantees the existence of an edge mode, constructed via the procedure of Section~\ref{Sec03a}.

After rotation, the Hamiltonian takes the form of Eq.~\ref{Eq15}, whose tight-binding chain is depicted in Fig.~\ref{Fig06}(b).  In this rotated frame, the two SSH chains acquire staggered on‐site energies
\begin{equation}\label{Eq31}
\epsilon'(k_x) = M_z(k_x)
= \bigl[M + \Delta(k_x)\bigr]\cos\theta \;+\; t'(k_x)\sin\theta,
\end{equation}
and are coupled by an interchain hopping
\begin{equation}\label{Eq32}
t_3(k_x) = -M_x(k_x)
= -\bigl[M + \Delta(k_x)\bigr]\sin\theta \;+\; t'(k_x)\cos\theta.
\end{equation}
Rearranging the atomic sites, as in Fig.~\ref{Fig06}(c), makes manifest the two decoupled 1D SSH chains with alternating hoppings $t_1$, $t_2$ and interchain link $t_3$, mirroring the structure studied at $k_x=\pi$ but now parametrically dependent on $k_x$.


\subsubsection{Topological Properties in the Absence of a Mass Term}

With $M=0$, Eqs.~\ref{Eq31}–\ref{Eq32} reduce to
$
M_z(k_x)=\Delta(k_x)\cos\theta + t'(k_x)\sin\theta,$ and
$
-M_x(k_x)=-\Delta(k_x)\sin\theta + t'(k_x)\cos\theta,
$
where $\Delta=2\lambda\sin k_x$ and $t'=2t\cos(k_x/2)$.  The winding number remains $\nu=1$ whenever $M_x(k_x)<\tilde t(k_x)$, signaling a robust edge state constructed as in Section~\ref{Sec03a} with appropriately adjusted amplitudes and normalization.  At the critical condition $M_x(k_x)=\tilde t(k_x)$, the $z$–axis touches the contour of $\tilde{\mathbf{d}}(k_y)$, one root of the recurrence reaches unity, and the edge‐state wave function diverges, marking the transition to $\nu=0$.  Beyond this point ($M_x>\tilde t$), no edge modes or topological signatures persist.

The energy dispersion of the $M=0$ edge mode follows directly from its role as an eigenstate of the rotated Hamiltonian in Eq.~\ref{Eq08}.  In particular, the on-site term 
\[
\epsilon'(k_x) = \Delta(k_x)\cos\theta + t'(k_x)\sin\theta
\]
acts as a $\sigma_z$–operator on the edge‐state subspace, yielding the energy eigenvalue
\begin{equation}\label{Eq34}
E(k_x) \;=\;\Delta(k_x)\cos\theta + t'(k_x)\sin\theta
= \frac{6t\lambda \sin(k_x)}{\tilde{t}(k_x)}
\end{equation}
where we have used $\sin\theta = 2t''/\tilde t$ and $\cos\theta = t/\tilde t$.  This analytic result agrees perfectly with our numerical band‐structure calculations represented in Fig~\ref{Fig01N} and the results reported in Ref.~\cite{Rahmati}.  

The band‐edge momentum $k_c$ is defined as the point in the Brillouin zone where the edge‐state dispersion merges with the bulk bands, signaling the boundary of the topological gap.  From the phase‐transition criterion 
\begin{equation}\label{Eq35}
M_z(k_c) = \pm \tilde{t}(k_c)\,,
\end{equation}
we substitute $M_z(k)= [M+\Delta(k)]\cos\theta - t'(k)\sin\theta$ (Eq.~\ref{Eq31}) and $\tilde{t}(k)$ from Eq.~\ref{Eq10}.  For $M=0$, this reduces to
\[
\Delta(k_c)\cos\theta = \pm \tilde{t}(k_c)\sin\theta
,
\]
so that
\begin{equation}\label{Eq37}
k_c = \pm \frac{2\pi}{3}\,.
\end{equation}
These analytic values for $k_c$ agree precisely with the numerical edge‐band crossings shown in Fig.~\ref{Fig01N}.

\subsubsection{Topological transition in the presence of a mass term}

To complete the analysis, we now consider the most general case in which a finite mass term $ M $ is included in the Hamiltonian. This allows us to explore the conditions under which a topological phase transition occurs based on the behavior of the edge states. According to the criterion for nontrivial topology, the winding number is non-zero when
\[
\left| M_x(k_x) \right| = \left| t'(k_x)\cos\theta - [M + \Delta(k_x)]\sin\theta \right| < \tilde{t}(k_x),
\]
in which case a stable edge state exists and can be constructed using the same procedure described earlier, with appropriate modifications to the on-site energies and normalization factor.

When the equality
\[
\left| M_x(k_x) \right| = \tilde{t}(k_x)
\]
is satisfied, the winding number drops to zero, indicating that the edge state merges with the bulk bands. This marks the occurrence of a topological phase transition. If instead
\[
\left| M_x(k_x) \right| > \tilde{t}(k_x),
\]
no edge states exist, and the system exhibits a trivial insulating phase. Thus, the condition for topological edge states in the presence of a mass term is encoded in the relative magnitude of the rotated mass component $ M_x(k_x) $ compared to $ \tilde{t}(k_x) $.

\vspace{1em}

To ensure the existence of topologically protected edge states, the edge-state dispersion must intersect zero energy. In the absence of a mass term, this intersection occurs at $ k_x = \pi $, as discussed previously. However, turning on a finite mass term shifts this crossing point. From Eq.~\ref{Eq31}, the energy of the edge state in the general case is given by
\begin{equation}\label{Eq38}
E(k_x) = [M + \Delta(k_x)]\cos\theta + t'(k_x)\sin\theta.
\end{equation}
Using the expressions for $ \cos\theta $ and $ \sin\theta $ in terms of hopping amplitudes, namely
\[
\cos\theta = \frac{t}{\tilde{t}(k_x)}, \qquad \sin\theta = \frac{t''(k_x)}{\tilde{t}(k_x)},
\]
we obtain the energy dispersion in the compact form
\begin{equation}\label{Eq39}
E(k_x) = \frac{1}{\tilde{t}(k_x)} \left[ t (M + \Delta(k_x)) + t''(k_x) t'(k_x) \right].
\end{equation}

Now, setting $ E(k_x) = 0 $ defines the momentum $ k_0 $ at which the edge-state band intersects zero energy. Solving this equation yields the critical value of the mass term for which this occurs:
\begin{equation}\label{Eq40}
M = \frac{t''(k_0) t'(k_0)}{t} - \Delta(k_0) = 2t''(k_0)\cos(k_0/2) - 2\lambda \sin k_0.
\end{equation}
This equation determines the critical momentum $ k_0 $ and the corresponding value of $ M $ at which the edge state touches zero energy. 

From this result, we observe that a zero-energy crossing exists only if there is a solution to Eq.~\ref{Eq40} in the domain $ k_x \in [-\pi, \pi] $. In particular, since the maximum value of $ \Delta(k_x) = 2\lambda $ and the maximum value of $ t'(k_x) = 2t $, it follows that the zero-energy crossing is guaranteed only when
\[
M < 3\sqrt{3}\lambda,
\]
which corresponds to the topological phase boundary of the bulk Haldane model. Therefore, for $ M < 3\sqrt{3}\lambda $, a solution $ E(k_x) = 0 $ exists, and the edge-state dispersion connects the conduction and valence bands, realizing a topologically nontrivial phase. At the critical point $ M = 3\sqrt{3}\lambda $, a phase transition occurs, and for $ M > 3\sqrt{3}\lambda $, the edge states no longer cross zero energy.

It is worth noting that even for $ M > \sqrt{t^2 + 4\lambda^2} $, edge-localized states can still exist. However, in this regime, the edge states are gapped and do not bridge the valence and conduction bands, thereby lacking the hallmark connectivity of topological edge modes.

\section{Summary and Conclusion}\label{Sec04}

In this work, we performed a detailed topological analysis of the Haldane model by explicitly constructing a dimensional reduction to a generalized one-dimensional SSH-type model. This reduction allowed us to reinterpret the two-dimensional Chern number in terms of a momentum-resolved one-dimensional winding number ($\nu$), which takes the value $ \pm 1 $ within specific ranges of $ k_x $, and vanishes elsewhere. This dimensional mapping not only provides an intuitive framework for understanding the topological properties of the Haldane model but also establishes a clear one-to-one correspondence between the 2D topological invariant (Chern number) and its 1D analogue (winding number).

Using this correspondence, we analytically derived the conditions under which the $\nu$ is non-zero, and subsequently calculated the edge-state wave functions and their energy dispersion relations both in the absence and presence of the mass term $ M $. The behavior of the $\nu$ in momentum space also enabled us to identify the band-edge momentum $ k_c $, at which the edge states merge into the bulk spectrum, signifying a topological phase transition.

Importantly, we showed that the edge states remain robust in a finite region of $ k_x $ even after the 2D topological phase transition occurs. However, these post-transition edge states do not bridge the conduction and valence bands, and thus lose their chiral character and do not contribute to transport—highlighting a subtle but important distinction between edge-state existence and topological protection.

Our findings build upon and significantly extend earlier work by our group~\cite{Rahmati}, where we derived the edge-state wavefunctions of the same system using a perturbative approach. While that study successfully characterized the edge-state structure, it lacked a topological framework to describe their origin. In contrast, the present analysis not only recovers those results without resorting to any perturbation theory but also clarifies the full topological origin and nature of the edge states by means of an exact analytical treatment based on the $\nu$ formalism.

Moreover, our approach closely relates to recent developments such as the study in Ref.~\cite{Ad8378} , which employed a similar dimensional-reduction framework to analyze topological characteristics of superconducting heterostructures. However, unlike our explicit mapping between the Haldane model and its SSH-type reduction, that study dealt with more complex Hamiltonians lacking a direct decomposition into one-dimensional chains with clearly defined topological invariants. Furthermore, our work goes beyond analogy by providing precise analytic expressions for edge-state energies, band-crossing points, and critical conditions for topological phase transitions—features that were not addressed in Ref.~\cite{Ad8378} .

In summary, this study establishes an analytically tractable and conceptually transparent route for understanding topological phases and edge-state dynamics in the Haldane model. Our results not only clarify previous findings but also offer a broader perspective for applying dimensional-reduction techniques to other 2D topological systems. We believe that these insights can inform future investigations into edge-state engineering, robust transport in nano-ribbons, and the design of materials with programmable topological features.

\appendix\section{Closed-form expressions for on-site amplitudes and normalization factor in edge State\label{AppA}}
In this appendix, we seek to derive the closed forms of the on-site amplitudes, $\alpha_n$ and $\beta_n$, 
as well as the normalization coefficient, $N$, as presented in Eq.~\ref{Eq29}. This derivation will 
utilize the initial conditions and the recurrence relation that characterize the scenario depicted 
in Fig.~\ref{Fig05}(c) for the generated edge state on the sites of the 1D~SSH chains. 
Referring to the figure mentioned, we can denote the on-site amplitudes at the sites where the edge 
mode is non-zero as $\tilde{\alpha}_0$, $\tilde{\alpha}_1$, $\tilde{\alpha}_2$,  and $\cdots$.
The initial conditions for the sequence are:
$$\tilde{\alpha}_0=0,\qquad \tilde{\alpha}_1=1,$$
and the recurrence relation governing the sequence is given by:
$$t_2 \tilde{\alpha}_n+t_3 \tilde{\alpha}_{n-1}+t_1 \tilde{\alpha}_{n-2}=0.$$
To solve this recurrence relation, we assume a solution of the form
$$\tilde{\alpha}_n=x^n.$$
Substituting this assumption into the recurrence relation yields:
$$t_2 x^{n}+t_3 x^{n-1}+t_1 x^n =0.$$
Dividing by $ r^{n-2}$ leads us to the characteristic equation:
$$x^2 - b x - c =0,$$
in which $b=- t_3/t_2$ and $c=-t_1/t_3$. The solutions to this quadratic characteristic equation can 
be found using the quadratic formula:
$$x_1={1\over 2}\big[b+\sqrt{b^2+4 c}\big],\qquad x_2={1\over 2}\big[b-\sqrt{b^2+4 c}\big].$$
The general solution to the recurrence relation is expressed as a linear combination of the powers 
of the roots $x_1$ and $x_2$:
$$\tilde{\alpha}_n = C_1 x_1^n + C_2 x_2^n.$$
We apply the initial conditions to determine the constants $C_1$ and $C_2$. From $\alpha_0=0$, we have:
$$C_1-C_2=0,$$
and from $\tilde{\alpha}_1=1$, we obtain:
$$C_1 x_1 + C_2 x_2  =1.$$
These two equations can be solved simultaneously to find $C_1$ and $C_2$ leading to:
$$C_1 = -C_2 ={1\over x_1-x_2} = {1\over\sqrt{b^2+4c}}.$$
By combining our findings for $C_1$ and $C_2$, we can express the closed form of $\tilde{\alpha}_n$ 
as follows:
$$\tilde{\alpha}_n = {\big[b+\sqrt{b^2+4 c}\big]^n - \big[b-\sqrt{b^2+4 c}\big]^n\over 2^n \sqrt{b^2+4 c}}$$
According to the notation used in Eq. 12 for the normalized edge state wave function, we have:
$$\alpha_n=\tilde{\alpha}_{2n-1},\qquad \beta_n=\tilde{\alpha}_{2n};\qquad n=1,~2,~3,~\cdots.$$
Additionally, based on the definition of the normalization factor, $N$, we have:
$$N=\Big[\sum_{n=0}^{\infty}\tilde{\alpha}_n^2\Big]^{-1/2}=\Big[{1\over (x_1 -x_2)^2} \sum_{n=0}^{\infty}(x_1^n-x_2^n)^2\Big]^{-1/2},$$
leading to
$$N=\sqrt{(1+c) \left[(1-c)^2-b^2\right]\over 1-c}.$$

\section*{Acknowledgment}
The first author  would like to acknowledge the office of
graduate studies at the University of Isfahan for their support and
research facilities. Also, M. A., acknowledges the
support received through the Abdus Salam (ICTP) short visit program.

\end{document}